[invited review]

# Optical diffraction tomography techniques for the study of cell pathophysiology

Kyoohyun Kim[1,2], Jonghee Yoon[1,2], Seungwoo Shin[1,2], SangYun Lee[1,2], Su-A Yang[3], and YongKeun Park[1,2,4]*

[1]Department of Physics, Korea Advanced Institute of Science and Technology (KAIST), Daejeon 34141, Republic of Korea; [2]KAIST Institute of Health Science and Technology, Daejeon 34141, Republic of Korea; [3]Department of Biological Sciences, KAIST, Daejeon 34141, Republic of Korea; [4]TOMOCUBE, Inc., Daejeon 34141, Republic of Korea

*Corresponding author: yk.park@kaist.ac.kr

**Abstract**

Three-dimensional imaging of biological cells is crucial for the investigation of cell biology, provide valuable information to reveal the mechanisms behind pathophysiology of cells and tissues. Recent advances in optical diffraction tomography (ODT) have demonstrated the potential for the study of various cells with its unique advantages of quantitative and label-free imaging capability. To provide insight on this rapidly growing field of research and to discuss its applications in biology and medicine, we present the summary of the ODT principle and highlight recent studies utilizing ODT with the emphasis on the applications to the pathophysiology of cells.

**1. Introduction**

Three-dimensional (3-D) optical microscopy techniques have been an invaluable tool in modern biological and medical sciences. They have been utilized to investigate mechanisms and dynamics of biological cells and to provide insights to aid in the diagnosis of diseases [1]. For recent half a century, we have witnessed the developments of various 3-D optical microscopy techniques and their applications in cell biology and medicine; from confocal microscopy [2] to multiphoton microscopy [3], and to super resolution nanoscopy [4].

Among them, optical diffraction tomography (ODT) has recently created and demonstrated a wealth of possibilities for 3-D imaging of live cells. ODT is an interferometric technique which measures the 3-D refractive index (RI) distribution of optically transparent samples such as biological cells. ODT is analogous to X-ray computed tomography (CT), except that it uses visible light instead of X-ray. Recently, ODT has gained significant interest, and the applications using ODT has rapidly expanded due to the following advantageous:

- ODT does not require the use of exogenous labeling agents or dyes, such as fluorescence protein, organic/inorganic dyes, and quantum dots. In ODT, RI, an intrinsic optical parameter of material, is utilized as an imaging contrast. ODT eliminates the complicated sample preparation processes and overcomes the limitations of labeling agents or dyes such as photobleaching or phototoxicity.

- In ODT, the measurements of RI provide quantitative imaging capability. The values of RI can be translated into various useful parameters such as protein concentrations and cellular dry mass.

- The instrumentational requirements for ODT are relatively simple and inexpensive. Unlike 3-D nonlinear optical microscopy techniques such as multiphoton microscopy or Raman microscopy, ODT utilizes low-power continuous-wave lasers for illumination and conventional image sensors such as CCD or CMOS.

ODT was first theoretically proposed in 1969 by E. Wolf [5] and followed by geometrical interpretation by Dändliker & Weiss [6]. From the late 70s, ODT has been experimentally demonstrated by early works [7-12]. Recently, there have been significant technical advances in ODT thanks to the advancements of laser sources, detecting devices, and computing powers. Also, the limitations and challenges of using labeling agents in 3-D cell imaging have become significant issues in several applications.

For these reasons, during the last few years, the field of ODT has started to expand to the various fields of applications, ranging from biophysics, cell biology, hematology, to infectious diseases.

Here, we summarize the recent advances on the developments of ODT techniques and discuss important applications of ODT in the fields of biology and medicine. This review is organized as follows: Section 2 briefly discusses the principles of ODT including optical instrumentations and reconstruction algorithms; Sections 3 summarizes the applications of ODT for the study of cell pathophysiology; Finally, perspectives for future developments and applications are discussed in Section 4.

## 2. Principles of ODT

In this section, the principles of ODT are discussed: various experimental configurations for ODT measurements and reconstruction algorithms are summarized. Further information on theoretical principles and experimental configurations for various ODT techniques can also be found elsewhere [13-16].

2.1 Optical setup for ODT

ODT solves an inverse problem of light scattering by a weakly scattering object. Typically, 3-D RI distribution of a weakly scattering sample or so-called a phase object is reconstructed from the measurements of multiple 2-D holograms of the object obtained with various illumination angles. It is analogous to X-ray CT (Fig. 1), in which 3-D absorption distribution of a human body is reconstructed via a filtered back projection algorithm, from the measurements of multiple 2-D X-ray images of the object obtained with various illumination angles. Both the ODT and X-ray CT shares the same governing equation – Helmholtz equation, and thus the principle of reconstruction algorithms is also identical.

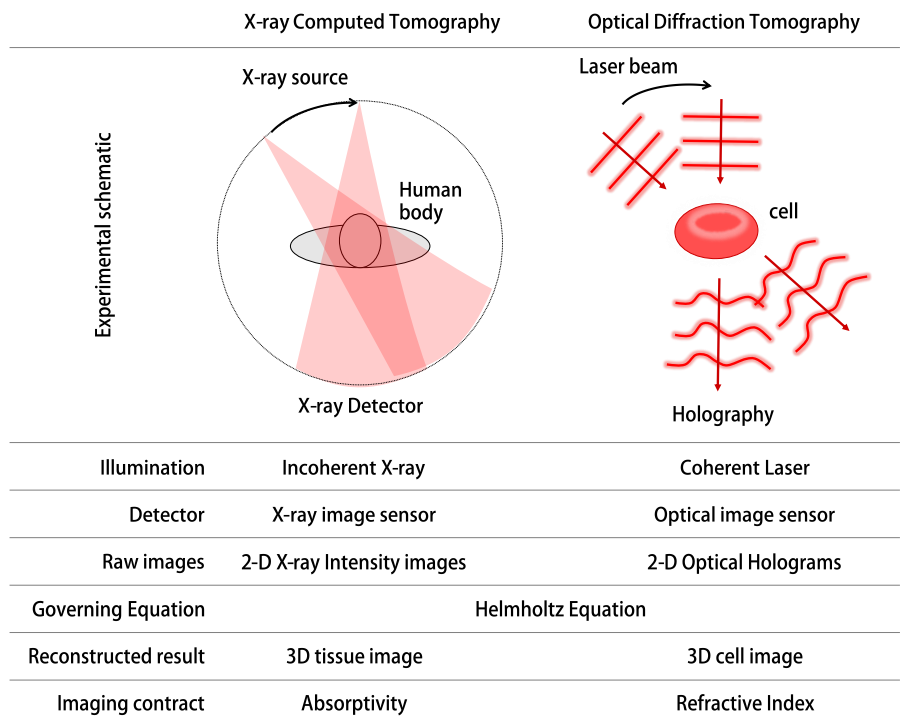

Figure 1. Comparison between X-ray CT and ODT.

Essentially, the optical setup for ODT consists of two parts: the illumination or sample modulation unit and the optical field recording unit. The optical field, composed of both the amplitude and phase information about a sample, is recorded employing the principle of interference. Diverse configurations are available for the optical field recording unit, including off-axis interferometry, phase shifting interferometry, or using transport of intensity equation. In-depth summaries on various 2-D field measurements techniques and field retrieval algorithms can be found in elsewhere [17-20].

To control illumination beams inpinging onto a sample, various techniques have been proposed, which can be classified into illumination scanning [21-24] and sample rotating schemes [25-27].

2.1.1 Illumination scanning ODT

For systematic controlling the angle of illumination beams, various types of beam rotators have been employed as depicted in the Figs. 1(a1–a3). Typically, galvanometer-based rotating mirrors have been used as a beam rotator, which controls the angle of the illumination beam impinging onto a sample by tilting mirrors of the galvanometer located at the conjugate plane of the sample [Fig. 1(a1)]. Due to the use of mirrors, beam scanning with a galvanometer can be executed with minimal laser power loss. The use of the galvanometer, however, is inherent in mechanical instability including position jittering induced by electric noise and positioning error at high voltage values caused by the nonlinear response. Moreover, due to the geometry of the dual-axis galvanometer, rotational surfaces for two independent axes cannot be exactly conjugated to the sample plane, which causes unwanted additional quadratic phase distribution on the illumination beam. To overcome this optical misalignment, placing two single-axis galvanometers at the separate conjugate plane relayed by additional 4-$f$ lenses can be used, but it inevitably requires a bulky optical system that may cause additional phase instability due to long optical paths.

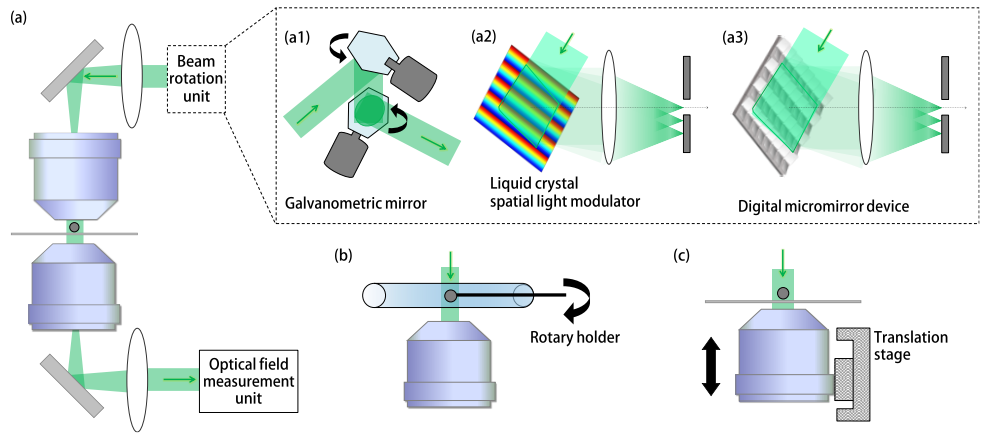

Figure 2. Experimental setups for standard ODT techniques. (a) Beam scanning methods using a beam rotator such as (a1) a dual-axis galvanometer mirror, (a2) a liquid crystal spatial light modulator, and (a3) a digital micromirror device. (b) A sample rotating method using a rotary holder, and (c) axial scanning of objective lens using a translational stage.

Recently, a spatial light modulator (SLM) using liquid crystals was utilized as a beam rotator [28]. To control the angle of the illumination beam, the SLM located at the conjugate plane to the sample displays wrapped linear phase ramp. Then, a plane wave with a desired propagating direction can be generated from the first-order diffracted beam while unwanted diffracted beams are blocked by spatial filtering [Fig. 1(a2)]. Since the use of the SLM does not contain any mechanically moving part, the ODT measurements can be performed with high stability. Also, the SLM can correct the wavefront distortion of an illumination beam to generate clearly plane waves at the sample plane, which enhances the accuracy of ODT measurements. However, due to the limited diffraction efficiency of the SLM and the spatial filtering, available illumination beam power significantly decreases. Furthermore, the slow innate response of liquid crystal realignment in the SLM slows down the scanning rate of the illumination beam, and expensive costs of an SLM may also limit wide applications in biological fields.

More recently, a digital micromirror device (DMD) was employed as a beam rotator for the ODT technique with high speed and stability [29, 30]. The DMD consists of hundreds of thousands of switchable micromirrors, which can be independently controlled between on and off states with ultra-high modulation speed reaching few tens of kHz. To steer the angle of the illumination beam, the DMD located at the conjugate plane to the sample displays a binary pattern, called the Lee hologram [31] to generate a plane wave with a desired illumination angle [Fig.1(a3)]. Similar to the use of an SLM, the illumination control with a DMD can be performed in high stability, and also able to correct wavefront distortion. Furthermore, the use of a DMD is cost efficient and enables fast illumination control due to the ultra-fast refreshing speed of a DMD. Because binary intensity holograms are projected, there is a limitation in the dynamic range of phase modulation and a beam size of the illumination with a DMD. Also, undesired diffraction from a DMD causes speckle noise and reduction in beam power.

2.1.2 Sample rotation ODT

Instead of illumination control, a sample of interest can be rotated, and the diffracted light fields at various rotation angles can be used for the reconstruction of 3-D RI map of the sample [Fig. 1(b)]. Contrast with the beam scanning method, a whole range of viewing angles can be addressed in the sample rotating method. To experimentally achieve the rotation of a sample, a rotary microcapillary is typically utilized [25-27]. Compared to illumination scanning ODT, which requires complicated optical setups for beam rotators, sample rotation ODT employs relatively simple optical setups, and reconstructed tomograms exhibit isotropic spatial resolution along the lateral and axial directions. Mechanical rotation of samples, however, limits data acquisition speed. Also, perturbation occurred during mechanical rotation, called radial run-out [32], and field distortion due to the refraction from the cylindrical microcapillary require additional numerical correcting algorithms [33]. More importantly, the sample rotation method can cause deformation of live biological cells because of the viscoelasticity of cells, possibly resulting in artifacts in reconstructed tomograms.

Recently, to avoid the perturbations and the refraction due to the use of the microcapillary, sample rotation methods exploiting optical trapping have been proposed [34-36]. Optical tweezers ware used to rotate RBCs in a microfluidic channel in order to measure 3-D morphology of cells [34]. Recently, optical force rotated a yeast cell [35] and a myeloid precursor cell [36] with precise rotation angles, from which 3-D RI distributions of rotating samples are reconstructed. Sample rotation with optical trapping, however, may induce morphological deformation of biological cells during rotation, and applications are limited to samples which can be floated; it would be challenging to measure optical adherent cells or soft cells using the optical trapping method.

2.1.3 Other methods

Alternatively, axial scanning of either an objective lens or a sample stage has also been utilized for reconstructing 3-D RI distribution of biological specimens [Fig. 1(c)] [37-39]. These methods can simplify optical setups, but exhibits limited axial resolutions. The use of low-coherence sources or the deconvolution has been reported for better axial resolutions [37-39]. Also, acquisition speed is also limited because mechanical scanning of an objective lens or a sample stage is slower than the use of galvanometric mirrors or DMDs. Also, scanning wavelengths of the illumination beam has also been utilized for measuring 3-D RI maps of samples [40, 41]. Recently, lens-free holographic and phytography techniques have been introduced which simplifies optical setup for ODT [42, 43].

2.2 Reconstruction procedures for ODT

In order to reconstruct the 3-D RI distribution of biological cells from measured multiple 2-D optical fields, several computational steps are required: (i) the application of appropriate reconstruction algorithms for solving an inverse scattering problem; (ii) the application of weak-scattering approximations depending on the scattering property of a sample, and (iii) regularization processes to fill information caused from limited measurements.

2.2.1 The reconstruction algorithms: projection and diffraction tomography algorithm.

The reconstruction algorithm can be classified into two groups: the diffraction algorithm and the projection algorithm. They are both inverse scattering solvers, but the difference is whether the effects of light diffraction is considered or not. Generally, ODT implies the reconstruction of 3-D RI maps using the diffraction algorithm. A few early works employed the projection algorithm for the study of cells [22]. However, it has been shown that individual biological cells exhibit significant light diffraction, especially for cells with complex internal structures [38].

To demonstrate that the use of the projection algorithm causes artifacts in measuring 3-D RI maps of microscopic samples, experiments were performed with a standard sample with known geometry. As shown in Fig. 2, a reconstructed tomogram of a polystyrene (PS) bead with the diameter of 10 μm placed at the focal plane exhibits uniform RI distribution while a PS bead at the defocused plane shows diffracted patterns, i.e. axially elongated shape.

The projection tomography assumes that each 2-D optical field image corresponds to the projection of the 3-D object at certain illumination angle and light diffraction inside samples is ignored. In other words, measured optical phase delay is assumed to be an integration of RI values along a straight light. In this case, the 2-D Fourier spectrum of the measured optical field at a certain illumination angle is identical to the inclined planar slice of 3-D Fourier spectra of the object, according to the Fourier slice theorem [13], as shown in Fig. 2(b). Thus, 3-D Fourier spectra can be mapped by a series of 2-D Fourier spectra which are obtained from the measured multiple 2-D optical fields with various incidence angles. Finally, the inverse Fourier transform of the mapped 3-D Fourier spectra provides the reconstruction of the 3-D RI distribution of the sample [Fig. 2(d)].

Implementation of the projection tomography algorithm is straightforward: applying inverse Radon transformation to the series of the measured optical fields with various incident angles, which is called as a sinogram, directly provides the reconstruction of a 3-D RI map. The projection algorithm is valid when the wavelength of illumination is much smaller than the size of a sample, thus, the wave propagation can be treated as projection. In X-ray CT, this condition is valid for general conditions, but this limits applications in the optical wavelengths. In the visible wavelengths, the sizes of the cellular components are compatible with the wavelengths of an illumination beam. Thus, light diffraction effects inside cells are not negligible so that the application of the projection algorithm deteriorates the image quality of the reconstructed tomograms, depending on samples. Although the projection algorithm has been successfully applied to the tomographic RI measurements of optical elements [25, 44] and biological samples [22, 45-47], it was shown that effects of light diffraction cannot be ignored even at the individual cell levels [38]. Thus, the use of the diffraction algorithm is recommended except for some special cases when the size of a sample is much large than the wavelength and RI contrast in a sample is negligible. Otherwise, the use of the projection algorithm may result in incorrect RI values and distorted shapes in the reconstructed tomograms. These unwanted artifacts become severe for objects located at defocused planes or for samples with a high optical phase delay.

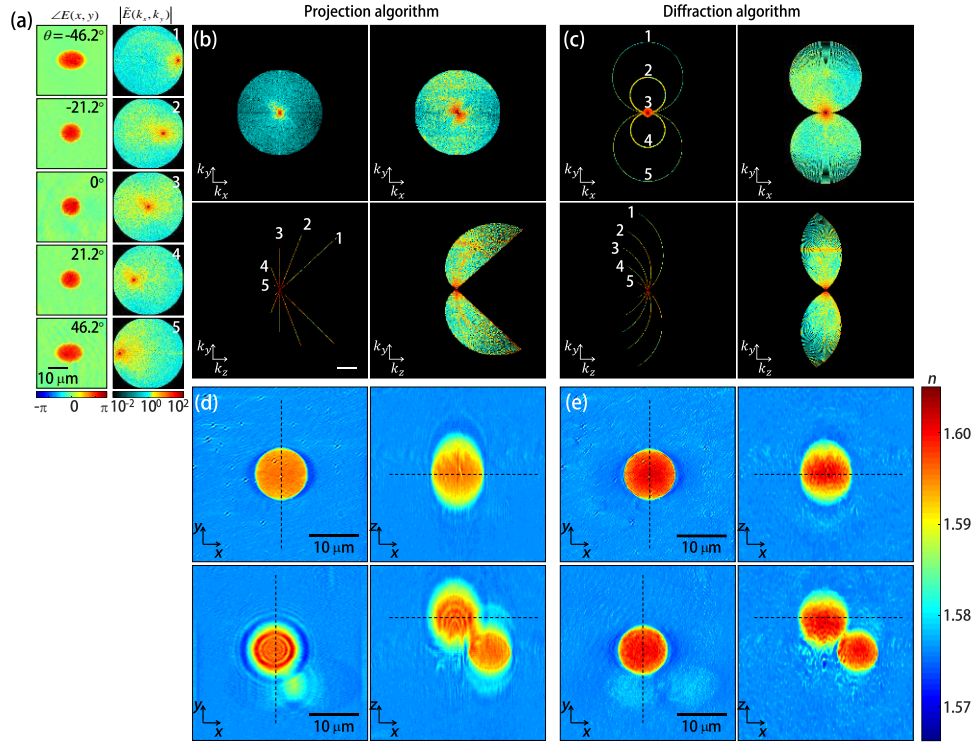

Figure 2. The principle of optical projection and diffraction tomography algorithm. (a) Quantitative phase images of a polystyrene bead with the diameter of 10 μm illuminated at various incidence angles, and corresponding 2-D Fourier spectra of complex optical fields. (b-c) Object functions in the 3-D Fourier space mapped by (b) projection tomography and (c) diffraction tomography. (d-e) Cross-sectional slices of the 3-D RI distribution of the polystyrene bead reconstructed by (d) projection tomography and (e) diffraction tomography.

The diffraction algorithm considers light diffraction inside a sample by solving the Helmholtz equation for the incident and scattered optical fields. Thus, the 3-D RI distribution of a sample can be obtained with more precise shapes and RI values of a sample than the case with the projection algorithm [5]. According to Fourier diffraction theorem, 2-D Fourier spectra of an optical field at a certain illumination angle is mapped onto the surface of a hemisphere called Ewald sphere. The center position of the Ewald sphere is translated from the origin in the 3-D Fourier space by the distance and direction corresponding to the ***k*** vector, or spatial frequency, of the incident angle [Fig. 2c]. Mapping Fourier spectra of multiple optical fields at various incident angles fill the 3-D Fourier space and 3-D inverse Fourier transform of the 3-D Fourier space results into the reconstruction of the 3-D RI distribution of a sample [Fig. 2e]. Tomograms reconstructed by this algorithm show more precise RI values and shapes of samples than the case with the projection algorithm, especially when optical fields of a sample are severely distorted either by highly scattering internal structures or when a target object is located at the defocused plane [Figs. 2d and 2e]. The detailed and rigorous mathematical derivations of the diffraction algorithm can be found elsewhere [5, 13, 21, 48]. Quantitative comparisons of tomogram reconstruction quality between the projection and the diffraction algorithm have been reported with various types of samples, including multimode fibers [49], colloidal particles [48], and malaria-infected human red blood cells (RBCs) [38].

2.2.3 Weak scattering approximation: Born vs. Rytov approximations

Finding the 3-D scattering potential, or the 3-D RI map, of a sample from the measured multiple 2-D optical fields, is an ill-defined inverse problem, which cannot be directly solved. However, under the first-order weak scattering approximation, the Helmholtz equation can be linearized, and the analytic solution for this inverse problem can be found. The weak scattering approximation for the linearization includes the Born or Rytov approximation [5, 13, 50].

The first-order Born approximation assumes that the scattered optical field from a sample is significantly weaker than the incident optical field, and assumes the total optical fields inside the sample as the incident optical field.

Specifically, the first-order Born approximation is valid when the total optical phase delay of a sample $\Delta\phi$ is smaller than $\pi/2$ [48]. However, the optical phase delay of most biological cells exceeds $\pi$ rad. For example, the cell thickness of 10 μm and the RI contrast between a cell and a medium greater than 0.03 are already beyond the valid regime of the first-order Born approximation. As shown in Figs. 3a and 3c, the Born approximation successfully reconstructs 3-D RI distribution of a 5-μm PS bead ($\Delta\phi = 0.41\ \pi$ rad.) whereas it fails to reconstruct for a 10-μm PS bead ($\Delta\phi = 0.83\ \pi$ rad.).

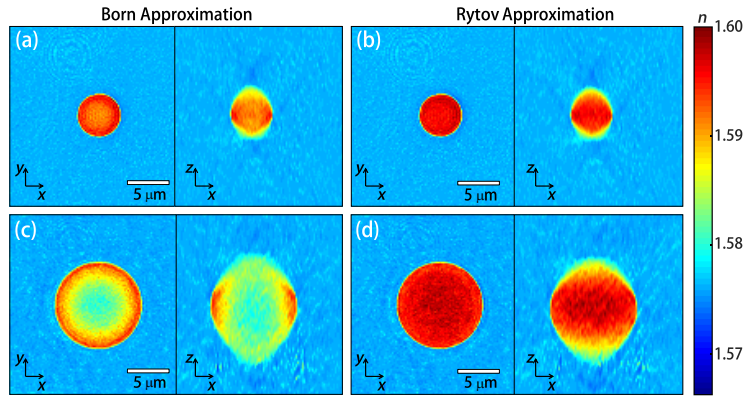

Figure 3. Comparison between (a & c) the first-order Born and (b & d) the first-order Rytov approximation in ODT. (a-b) Cross-sectional slice of the 3-D RI distribution of a 5-μm-diameter polystyrene bead with (a) Born and (b) Rytov approximation. (c-d) Same as (a-b) for a 10-μm-diameter polystyrene bead.

In contrast, the first-order Rytov approximation considers the total optical field as a complex amplitude, and assumes that the spatial variations of the complex amplitude in the wavelength scales are smaller than 1 [51]. For that reason, the validity of the first-order Rytov approximation is independent of the total optical phase delay as:

$$\Delta n \gg \left(\frac{\lambda \nabla \varphi_s}{2\pi}\right)^2,$$

where $\Delta n$ is the RI contrast between a cell and a medium and $\nabla \varphi_s$ is the spatial variations of the complex amplitude. The above relation can be deduced to $\Delta n \ll n_m$ for the weak scattering sample where $\nabla \varphi_s$ is linearly proportional to $\Delta n$ [52]. As shown in Figs. 3b and 3d, the first-order Rytov approximation remains valid whenever the spatial gradient of the optical phase is not significant. Because most biological cells have a small spatial gradient of RI distribution, especially when cells are submerged in a medium, the first-order Rytov approximation is suitable for the tomographic reconstruction.

2.2.4 Regularization process

Unlike X-ray CT where full 360° angle of illumination is possible, the ODT techniques employing the beam scanning method has a limited angle range of illumination, which is mainly determined by the numerical aperture of objective lenses. Consequently, in the Fourier space, there exists the inaccessible region corresponding to the spatial frequencies beyond the maximum spatial frequencies of the imaging system. This issue due to the inaccessible region is called as the missing cone problem because the shape of the inaccessible region resembles a cone. This missing cone problem causes shape elongation and under-estimated RI values in reconstructed tomograms [53, 54]. Although the missing cone problem is unavoidable for the ODT techniques with the beam scanning, it can also occur in the cases with the sample rotation method; incomplete rotation of a sample results in a similar situation comparable to the limited numerical aperture in the beam scanning method. Rotating a sample around one axis also generates a missing part in the Fourier space which shape resembles an apple core [55].

To relieve the missing cone problem and enhance the quality of tomogram reconstruction, several iterative algorithms have been proposed to fill the unassessed information based on *a priori* information about an object such as the non-negativity constraint algorithm, the edge-preserving regularization, and the total-variation regularization. Figure 4 shows comparisons of the 3-D RI distribution of a hepatocyte reconstructed without the regularization process [Fig. 4(a)], which is reproduced from Ref. [56]. For comparison, the 3-D RI maps of the hepatocyte obtained using the various regularization methods including the non-negativity constraint, the edge-preserving regularization, and the total-variation regularization algorithm are presented in Fig. 4(b)–(d), respectively. Detailed information about these iterative algorithms and comparative studies can be found elsewhere [56, 57].

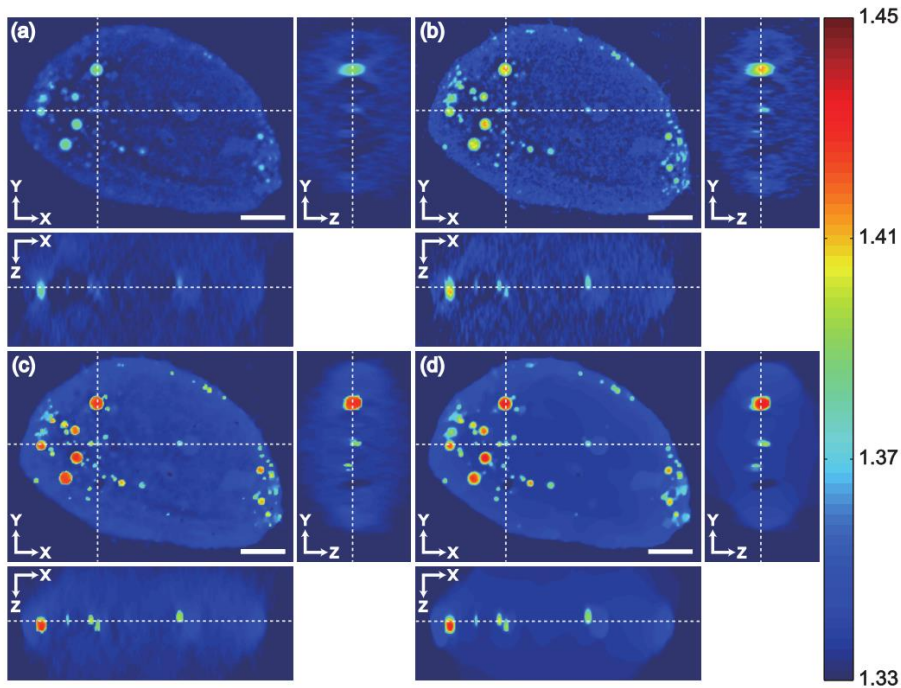

Figure 4. Comparison of 3-D RI distribution of a hepatocyte reconstructed by (a) direct Fourier transform, (b) non-negativity constraint, (c) edge-preserving regularization, and (d) total-variation regularization algorithm. Reprinted from Ref. [56] with permission.

Recent works have shown that this missing cone problem can be overcome alternatively. Complex deconvolution using a predefined coherent transfer function (CTF) corrects aberration induced by the missing cone problem and achieves sub-100-nm lateral resolution using objective lenses with high numerical aperture [23]. The missing cone problem can also be relieved using a machine-learning algorithm based on the beam propagation method [58].

2.2.5. Spatial Resolution in ODT

Theoretically, the spatial resolution of ODT is defined as the maximum spatial frequency in reconstructed Fourier spectra. This maximum spatial frequency is mainly determined by the wavelength of light $\lambda$, the numerical aperture of an imaging system $NA$, the RI value of immersion media $n_i$, and the optical configuration for ODT [21], which is summarized in Table 1.

|  | Lateral resolution | Axial resolution |
|---|---|---|
| The sample rotation method | $\dfrac{\lambda}{2NA}$ | $\dfrac{\lambda}{2NA}$ |
| The beam scanning method | $\dfrac{\lambda}{4NA}$ | $\dfrac{\lambda}{2}\dfrac{1}{n_i - \sqrt{n_i^2 - NA^2}}$ |

Table 1. Lateral and axial resolution in ODT.

The axial resolution of illumination-scanning ODT is very sensitive to *NA* and $n_i$. For instance, with a 532-nm-wavelength laser, an objective lens with NA = 0.9 provides the axial resolution of 472 nm, while and an objective lens with NA = 0.45 gives the axial resolution of 2.49 μm. In a recent work, it was theoretically proposed that ODT can provide isotropic lateral and axial resolution of $\lambda/(4NA)$ by combining sample rotation and illumination scanning [59].

## 3. Study of cell pathophysiology using ODT

Due to the aforementioned advantages, there is an increasing number of studies employing ODT for the study of cell pathophysiology. Label-free visualization of live cells has a potential for the investigation of functions and mechanisms at the individual cell level. The following section summarizes physiological parameters of cells accessible with ODT and then highlights the representative pathophysiological study of cells with ODT techniques.

3.1 Quantitative physiological parameters of cells accessible with ODT

Quantitative 3-D imaging capability of ODT enables the retrieval of various physiological parameters, including morphological and biochemical parameters.

3-D RI maps of a cell directly provide several morphological parameters. First, from distinct boundaries in a 3-D RI map provide information about the volume *V* and surface area *A* of cells and subcellular organelles. From the retrieved values of *V* and *A*, sphericity $\psi$, a dimensionless parameter indicating how much an object is spherical, can be calculated as $\psi = \pi^{1/3}(6V)^{2/3}/A$. To address the cellular or subcellular boundaries from measured 3-D RI maps, isosurfaces can be extracted [38]. Recently, it was shown that these boundaries can be effectively retrieved using the multidimensional transfer function, in which both the spatial gradient of RI and the RI values are simultaneously utilized [30]. These morphological features can be potentially valuable for investigating structural characteristics of cells and effects of internal and external stimuli on cellular functions such as osmotic stress and drug treatment.

The values of RIs can provide biochemical information about cells. For example, the values of RI can be translated into the local concentration in the cell cytoplasm. Because the RI values of a solution are linearly proportional to the concentration of solutes, the RI values of cell cytoplasm can be quantitatively converted into the local cytoplasmic concentration. Measured RI value of a biological sample *n* is linearly proportional to dry mass density *C* as $n = n_m + \alpha C$, where $n_m$ is the RI value of a medium surrounding a cell, $\alpha$ is RI increment [60, 61]. Furthermore, together with the retrieved value of a cell volume, the dry mass – the total mass of non-aqueous materials inside a cell – can also be calculated from the concentration, because mass is a product of concentration and volume. The values of $\alpha$ are similar regardless of protein types; most amino acids exhibit very similar values of $\alpha$ [62-65]. Because biological cells mostly consist of proteins (amounts of carbohydrates, lipids, and nucleic acids can be ignored), the RI increment for cells can be assumed as a constant when calculating the dry mass and the dry mass density of biological cells.

Recently, there have been several studies using ODT to investigate the physiology of biological cells, including red blood cell (RBC) [29, 30, 38, 66-71], white blood cell (WBC) [27, 30, 72, 73], and various eukaryotes [27, 30, 48, 67, 74-78]. The following sections highlight representative work.

3.2 Studies with human red blood cells

Human RBCs or erythrocytes have a donut-like shape with a diameter of 8.5 μm and a thickness of 2.5 μm, whose function is to carry oxygen from lungs to tissues. RBCs exhibit remarkable deformability, which enables them to pass through small capillaries even narrower than RBC sizes. Physiologically relevant cell indices for RBCs include cell volume, surface area, sphericity, Hemoglobin (Hb) content, and Hb concentration. In the clinic, some of these RBC indices are routinely measured with a complete blood count (CBC) machine which exploits various analytical methods such as light scattering, electrical impedance, and fluorescence.

Employing ODT, these RBC indices can be quantitatively and precisely measured at the individual cell level, whereas conventional CBC machines measure averaged values from many cells [66]. Morphological parameters such as cell volume, surface area, and sphericity are retrieved from 3D RI maps of RBCs. Cytoplasmic Hb concentration is directly converted from measured RI values, and Hb contents can be calculated from Hb concentration and cell volume because RBC cytoplasm mainly consists of Hb solution without a nucleus or other subcellular organelles. Several pathophysiological states in RBCs associated with morphology have been investigated using ODT, including iron-deficiency anemia and hereditary spherocytosis [66]. Recently, ODT was employed for the characterization of RBCs from cord blood of newborn infants [68]. This study reported that cord RBCs exhibit smaller cell surface area and volume, but the elevated level of Hb contents, compared to adult RBCs.

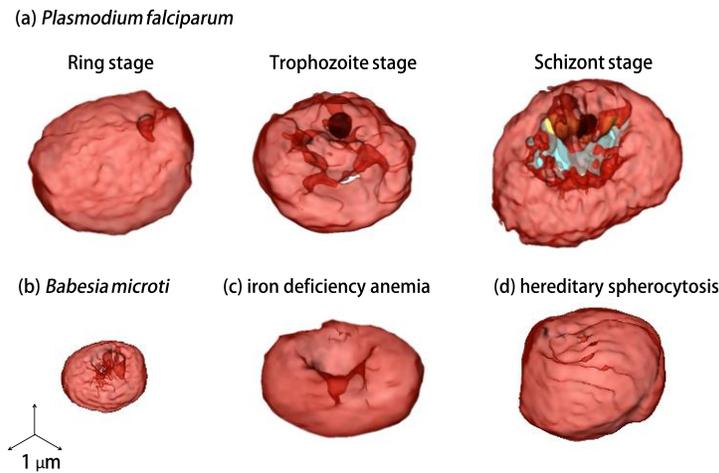

Figure 6. 3-D isosurfaces of representative RBCs under various pathological conditions: (a) RBCs parasitized by *Plasmodium falciparum* at the ring (left), trophozoite (center) and schizont (right) stage, respectively. (b) Mouse RBC parasitized by *Babesia microti*. (c) RBC from a patient with iron deficiency anemia. (d) RBC with a patient with hereditary spherocytosis. Arrows indicate 1 μm along the *x*, *y*, and *z*-axis.

Recently, Lee *et al*. investigated the *in-vitro* effects of ethanol on morphological and biochemical properties of human RBCs using ODT [71]. In this work, no significant alterations have been observed in shapes and Hb contents of RBCs, whereas significant enhancements in dynamic membrane fluctuation have been measured which indicates enhanced deformability and membrane fluidity [71]. More recently, properties of individual RBCs in stored blood and the effects of a blood preservation solution, CPDA-1, have been systematically investigated by measuring 3D RI maps of individual RBCs with various blood storage durations and conditions [70].

3.3 The study of white blood cells

WBCs are classified into several cell types including B cells, T cells, and macrophages; their roles are defending the host against harmful pathogens, abnormal cells, and other invaders. It is well known that WBCs shows alterations in morphological and biochemical characteristics in response to immune stimuli, such as bacterial infection and sepsis [79].

ODT has been applied for label-free quantitative analysis of several WBCs and demonstrated potential for the study of WBCs. Yoon *et al*. have shown the characterization of morphological and biochemical properties of individual lymphocytes. In particular, alterations in these parameters of WBCs in response to immune stimulus and internalization of a microsphere via phagocytosis have also been investigated using ODT [27, 72]. Recently, ODT in combination with the optical tweezers technique has shown potential for the investigation of the cellular response of macrophages to external mechanical stimuli [73]. When a microscopic particle, trapped and optically manipulated with the optical tweezers, was approached to a macrophage, the deformation of the cell upon this mechanical stimulus has been precisely visualized from the measured 3-D RI maps over time.

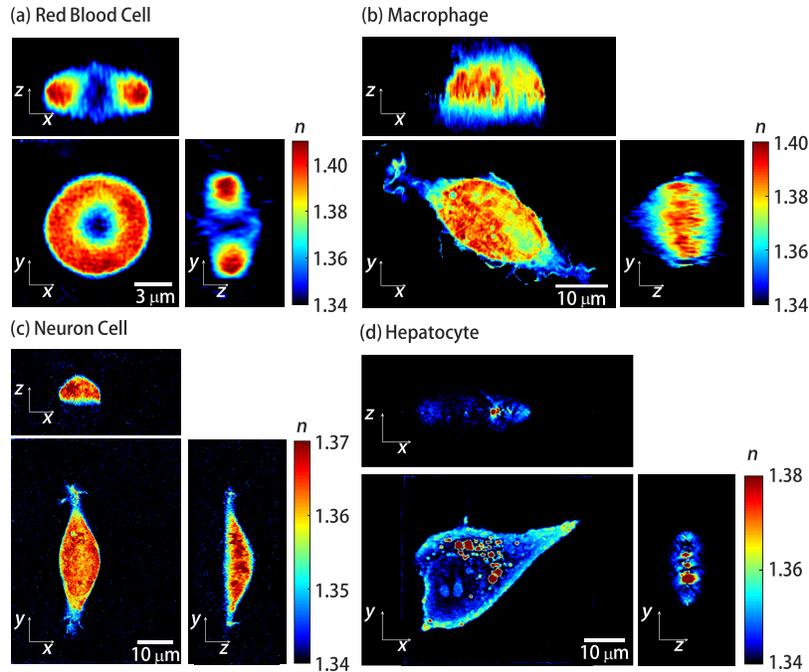

Figure 5. Representative 3-D RI tomograms of (a) RBC, (b) macrophage, (c) hepatocyte, (d) neuron, (e) phytoplankton, and (f) downy hair. Color bar indicates refractive index values.

3.4 Physiology of other eukaryotic cells

The applications of ODT for the study of eukaryotic cells are in the beginning stage, and only a few recent works have been reported. Nonetheless, ODT has shown great potential for the visualization of eukaryotic cells, because it provides not only physical and biochemical properties of eukaryotes, it can also effectively visualize intracellular organelles from the measured 3-D RI map. It has been shown that ODT is capable of visualizing plasma membrane, nucleus, nucleoli, chromosomes, and internal vesicles [27, 74, 80, 81]. Using dual wavelength ODT using visible and ultraviolet wavelengths, Sung *et al*. has shown that the dry mass of condensed chromosomes, as well as entire cells, can be measured and quantified with ODT, by measuring the 3-D RI maps of HT-29 and T84 cells in metaphase [75]. In particular, this work also presented that the use of the diffraction algorithm with the regularization process provides more precise measurements in compared to one obtained with the projection algorithm, which is confirmed by the 3-D image obtained by laser scanning confocal fluorescence microscopy. Kuś *et al*. have measured the 3-D RI maps and analyzed the morphological parameters of U937 and HT-1080 cells using the sample rotation method [27]. Utilizing holographic optical tweezers for sample rotation, Habaza *et al*. has measured the 3-D RI distribution of yeast and validate measured internal structures from 3-D RI information with confocal fluorescent microscopy [35].

Measuring RI values in live eukaryotes enables to investigate cellular responses according to external environmental changes. Fang-yen *et al*. reported rapid increases in RI values of cytoplasm upon exposure to the acetic acid solution, while there were no significant changes in nucleoli [82]. Also, 3-D dynamics of RI distributions of lipid droplets in human hepatocytes (Huh-7) has been characterized to investigate structural and biochemical changes of internal vesicles under chemical treatment [76]. 3-D shapes and internal structures of *E. coli*, prokaryotic cells and dynamic 3D RI images of synaptic network were visualized using ODT with the complex deconvolution technique [23].

3.5 Parasitology: malaria and babesiosis

Malaria is an infectious disease causing significant mortality. Transmitted to humans via mosquitos, malaria-inducing parasites first invades hepatocytes, and then RBCs. Imaging and visualization techniques for the study of malaria is crucial because infectious mechanisms regarding malaria are not entirely understood [83]. ODT has shown great potential for the study of malaria because it can effectively visualize host RBCs, invaded parasites, as well as other parasite-induced vacuole structures. Also, the label-free capability of ODT allows direct and easy visualization of malaria infected RBCs.

The 3-D RI tomograms of RBCs parasitized by *Plasmodium falciparum* (*pf*-RBCs) was measured with the optical projection algorithm by Park *et al*. [46]. In this work, intracellular parasites and formed hemozoin crystals were clearly measured in the reconstructed 3-D structures of *pf*-RBCs. Also, the decreases in Hb concentration in *pf*-RBCs due to the consumption of parasites were quantified. Recently, it was demonstrated that the diffraction algorithm provides better imaging quality for the reconstruction of *pf*-RBCs, in compared to the projection algorithm [38]. Reconstructed 3-D shapes of *pf*-RBCs in different stages of infection and hemozoin were shown in Fig. 6(a).

ODT has been successfully used for the study of babesiosis [69], an infectious disease transferring from animals to humans and showing similar to malaria, such as anemia and fever [84, 85]. Emergency human babesiosis is caused by an obligate parasite *Babesia microti* (*B. microti*), an intraerythrocytic protozoan of the genus *Babesia* [86, 87]. *B. microti* mainly invades RBCs and causes significant alteration in host RBCs.

Conventional techniques for imaging *B. microti*-invaded RBCs (*Bm*-RBCs) are based on optical microscopy with Giemsa-staining [86, 87]. However, these only provide limited structural information: parasitized vacuoles are not labeled with Giemsa and the discrimination of parasites rely on the experience of professional technicians. Park *et al*. has shown that measuring 3D RI maps offers a novel approach to the investigation of *Bm*-RBCs [69]. The morphological, chemical and mechanical parameters of *Bm*-RBCs are quantified at the individual cell level, which reveals significant changes resulting from *B. microti* parasitic invasion [Fig. 6(b)].

3.6 Emerging applications: virus infection and cancer biology

One of the emerging applications of ODT is the study of virus infection. Although individual viruses are too small to be optically resolved, ODT can effectively visualize structures of individual cells infected with the virus. Using an ODT system combined with confocal microscopy, Simon *et al*. have measured 3-D RI distribution and corresponding fluorescence images of human alveolar epithelial A549 cells infected with H3N2 influenza viruses [88].

Another interesting field of application is cancer biology. WC Hsu *et al*. has measured 3-D RI tomograms of cancerous and normal epithelial cells, and investigated backscattering light properties by applying the finite-difference-time-domain method to experimentally determined 3-D RI tomograms [89].

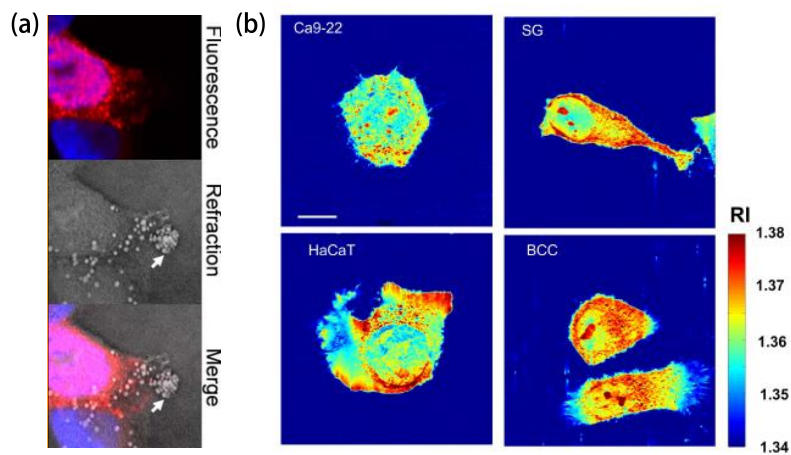

Figure 7. 3-D RI distribution of eukaryotic cells in various pathological conditions. (a) Multimodal images of human epithelial A549 cells infected with H3N2 influenza viruses. Red indicates viral nucleoprotein NP immunostaining. Blue indicates nucleus DAPI staining. Reprinted from Ref. [88] with permission. (b) *x* - *y* cross-sectional slices of 3-D RI tomograms of cancerous CA9-22 and BCC cell lines and two normal HaCaT and SG cell lines. Modified from Ref. [89] with permission.

## 4. Concluding remarks

In this review, the principles of ODT and their applications to the study of biological cells have been presented. The research work reviewed here and recent trends suggest that measuring 3-D RI maps of live cells may play a significant role in enhancing the understanding of cell pathophysiology. In particular, ODT will be widely used in biological and medical labs, especially when quantitative imaging capability and label-free imaging of cells are crucial such as immunotherapy using WBCs and stem cell research. However, this does not imply the ODT replaces existing optical microscopy techniques such as fluorescence confocal or multiphoton microscopy; ODT man is more efficiently used when complementary to existing cell imaging procedures.

The uses of ODT for studying cell pathophysiology have not yet been fully explored; indeed, this emerging field is at its very early stage with a few significant applications in a few fields. Considering the advantageous and potential of ODT, significant development, and wide application is expected. However, there are still many important issues in the study of cell pathophysiology using ODT that can be tackled by the good development and use of novel techniques. In particular, new techniques are needed to simplify new ODT systems. As mentioned in this review, active illumination schemes using a DMD, an SLM, or an LED array have demonstrated merits for practical applications of ODT to biological field [28, 29, 90]. The interferometric imaging techniques can also be further advanced for simple and stable measurements of optical fields. Recently reported techniques on common-path interferometric units have shown progressive advances in the interferometric imaging [24, 67, 91, 92]. Also, most of previous ODT techniques have used single visible wavelengths. The use of multiple wavelengths in ODT for hyperspectral RI tomography has been recently demonstrated [77, 93], which may provide molecular specificity for some special applications [94]. More importantly, the study of cell pathophysiology using ODT will be further accelerated when biologists and medical doctors extensively utilized this technique. Very recently, ODT is commercially available by spin-off companies [95], and this will further expedite the applications of ODT.

Also, the tomographic reconstruction algorithm can be further developed. For example, current existing reconstruction algorithms are essentially based on the first order approximation and thus are not applicable to thick samples exhibiting significant multiple light scattering. To further expand the applicability of ODT from single cells to multi-layered cells, new algorithms to take into account multiple scattering should be developed, although it will be challenging due to extremely high level of computational demands. It was noteworthy that new tomographic reconstruction algorithm based on machine learning was developed [58], which will bring new insights to the problem of multiple scattering. It is noteworthy that the experimental techniques used in ODT share the same principles with the methods used in measuring transmission matrices (TM) of highly turbid media [96-98]. TM relates the scattered light fields from complex media with indecent fields, which have been exploited in focusing and imaging through the complex media [99-102]. With the development of a clever reconstruction algorithm, the 3-D RI distributions of highly complex samples such as biological tissues may be directly measured with ODT.

Furthermore, regularization methods to alleviate the missing cone issues can also be advanced. Most of the conventional regularization methods are developed for the application of X-ray CT, and thus not optimized for the case of biological cells. In addition, the reconstruction speed in ODT should be enhanced for practical applications in biological and clinical settings. Recently, the use of GPU was reported to expedite the tomographic reconstruction processes in ODT [103].

Although this review article focused on the applications to the cell pathophysiology, the use of ODT can be further applied to various related research fields. For example, imaging tissue slices will provide detailed subcellular structures without labeling [104] or enable access to morphological features which is difficult to be imaged with conventional labeling agents. Also, as recently demonstrated, ODT can also be exploited for the study of phytoplankton [105] and hair [106] as well as industrial applications [49, 107].

We have witnessed that recent developments in optical imaging techniques have been effectively transferred from laboratories to clinics due to close interdisciplinary collaboration: successful examples include optical coherence tomography, multiphoton microscopy, and super-resolution nanoscopy. To successfully address critical issues in cell pathophysiology with ODT, it is crucial to develop interdisciplinary collaborations among clinicians, biologists, engineers and physicists. Considering the recent exponential growth of the field, we are optimistic that ODT will play important roles in various cellular studies.

**Acknowledgement**

This work was supported by KAIST, and the National Research Foundation of Korea (2015R1A3A2066550, 2014K1A3A1A09063027, 2012-M3C1A1-048860, 2014M3C1A3052537) and Innopolis Foundation (A2015DD126).

**Competing financial interests**

YongKeun Park is a co-founder and CTO of TomoCube, Inc.